\documentclass[twocolumn,showpacs,preprintnumbers,amsmath,amssymb]{revtex4}
\usepackage{graphicx}% Include figure files
\usepackage{dcolumn}% Align table columns on decimal point
\usepackage{bm}% bold math

\begin{document}

\preprint{APS/123-QED}

\title{A quantum key distribution and identification protocol \\ based on entanglement swapping}% Force line breaks with \\

\author{Fei Gao$^{1,2}$ \quad Fenzhuo Guo$^{1}$ \quad Qiaoyan Wen$^{1}$ and Fuchen Zhu$^{3}$\\
        (1. School of Science, Beijing University of Posts and Telecommunications, Beijing, 100876, China) \\
        (2. State Key Laboratory of Integrated Services Network, Xidian University, Xi'an, 710071, China)\\
        (3. National Laboratory for Modern Communications, P.O.Box 810, Chengdu, 610041, China)\\ Email: hzpe@sohu.com}

\date{\today}% It is always \today, today,

\begin{abstract}
A quantum key distribution and identification protocol is
proposed, which is based on entanglement swapping. Through
choosing particles by twos from the sequence and performing Bell
measurements, two communicators can detect eavesdropping, identify
each other and obtain the secure key according to the measurement
results. Because the two particles measured together are selected
out randomly, we need neither alternative measurements nor
rotation of the Bell states. Furthermore, less Bell measurements
are needed in our protocol than in the previous similar ones.
\end{abstract}

\maketitle

\section{\label{sec:level1} Introduction}

As a kind of important resource, entanglement \cite {EPR} is
widely used in the research of quantum information, including
quantum communication, quantum cryptography and quantum
computation. Entanglement swapping \cite{ZZHE}, abbreviated by ES,
is a nice property of entanglement. That is, by appropriate Bell
measurements entanglement can be swapped between different
particles. For example, consider two pairs of particles in the
state of
$|\Phi^+\rangle$,equivalently,$|\Phi^+\rangle_{12}=|\Phi^+\rangle_{34}=1/\sqrt{2}(|00\rangle+|11\rangle)$,
where the subscripts denote different particles. If we make a Bell
measurement on 1 and 3, they will be entangled to one of the Bell
states. Simultaneously, 2 and 4 will be also projected onto a
corresponding Bell state. We can find the possible results through
the following process:
\begin{eqnarray}
|\Phi^+\rangle_{12}\otimes|\Phi^+\rangle_{34}=\frac{1}{2}(|00\rangle+|11\rangle)_{12}
\otimes(|00\rangle+|11\rangle)_{34}\nonumber\\
=\frac{1}{2}(|0000\rangle+|0101\rangle +|1010\rangle
+|1111\rangle)_{1324}\nonumber\\
=\frac{1}{2}(|\Phi^+\Phi^+\rangle+|\Phi^-\Phi^-\rangle
+|\Psi^+\Psi^+\rangle+|\Psi^-\Psi^-\rangle)_{1324}
\end{eqnarray}
It can be seen that there are four possible results:
$|\Phi^+\rangle_{13}|\Phi^+\rangle_{24}$,
$|\Phi^-\rangle_{13}|\Phi^-\rangle_{24}$,
$|\Psi^+\rangle_{13}|\Psi^+\rangle_{24}$ and
$|\Psi^-\rangle_{13}|\Psi^-\rangle_{24}$. Furthermore, these
results appear with equal probability, that is,$1/4$. For further
discussion about ES, see \cite {ZHWZ,BVK,KBB,PBWZ}.

Quantum cryptography is the combination of quantum mechanics and
cryptography. It employs fundamental theory in quantum mechanics
to obtain unconditional security. Quantum key distribution (QKD)
is an important research direction in quantum cryptography.
Bennett and Brassard came up with the first QKD protocol (BB84
protocol) in 1984 \cite{BB84}. Afterwards, many protocols were
presented \cite
{E91,B92,BW92,GV95,HIGM,KI97,B98,LL02,PBTB,XLG,LCA}. Recently,
several QKD schemes based on ES were proposed
\cite{C2000,ZLG,C2001,D2004,CQ-PH,LLKO,ZYCP}. In
\cite{C2000,ZLG,C2001} the author introduced a protocol without
alternative measurements. It was simplified \cite{D2004} and
generalized \cite{CQ-PH} before long, and its security was proved
in \cite{LLKO}. Besides, Zhao \textit{et al}. presented a protocol
using ES on doubly entangled photon pairs \cite{ZYCP}.

In this paper we propose a QKD protocol based on ES, which needs
neither alternative measurements \cite{ZYCP} nor rotation of the
Bell states \cite{C2001,D2004,CQ-PH}. Simultaneously, we also use
ES to identify the legal users. The security against the attack
discussed in \cite{ZLG} is assured by certain classical means. See
Sec.II for the details of this protocol. The security is analyzed
in Sec.III and a conclusion is given in Sec.IV.
\section{The QKD and identification protocol}
Suppose the two legal communicators, Alice and Bob, share a bit
string $ID$, which is used to identify each other. The initial
$ID$ can be obtained by the means discussed in \cite{ZZ}. The
particular process of this scheme is as follows:

1. Prepare the particles. Alice generates a sequence of EPR pairs
in the state
$|\Phi^+\rangle_{AB}=1/\sqrt{2}(|00\rangle+|11\rangle)$.For each
pair, Alice stores one particle and sends the other to Bob.

2. Detect eavesdropping.

(1) Having received all the particles from Alice, Bob randomly
selects a set of particles out and makes Bell measurements on them
by twos.

(2) Bob tells Alice the sequence numbers and measurement results
of the pairs he measured.

(3) According to the sequence numbers, Alice performs Bell
measurements on the corresponding pairs, and compares her results
with Bob's. For example, consider one of the pairs Bob measured,
in which the sequence numbers of the two particles are $m$ and
$n$. Then Alice measures her $m$-th and $n$-th particles in Bell
basis, and compares the two results. As discussed in Sec.I, if
these particles were not eavesdropped, Alice and Bob should obtain
the same results. Otherwise there must be an eavesdropper (Eve) in
the channel, Alice will abort this communication.

3. Identify the users.

(1) Bob randomly selects a set of particles out from his left
sequence, and divides them into two subsets averagely, which are
denoted as $S_1$ and $S_2$, respectively.

(2) Bob makes Bell measurements on the particles in $S_1$ by twos.
Note that for each pair there are two sequence numbers (because
each particle has a sequence number) and one corresponding
measurement result. Here two notations, $P_1$  and $R_1$ , are
introduced to denote all the sequence numbers and all the
corresponding results, respectively. By the same means, Bob
measures the particles in $S_2$ . Similarly, the sequence numbers
are denoted as $P_2$ and the corresponding results as $R_2$ .

(3) Taking the shared information $ID$ as the key, Bob encrypts
$P_1$, $R_1$ and $P_2$  with the one-time-pad cipher:
$y=E_{ID}(P_1,R_1,P_2)$ , and sends $y$ to Alice through the
classical channel.

(4) When Alice received the ciphertext $y$, she decrypts it with
the help of $ID$: $P_1',R_1',P_2'=E_{ID}^{-1}(y)$ . According to
$P_1'$ , Alice measures the corresponding particles in Bell basis
and compares her results with $R_1'$ . If both results coincide,
Alice considers Bob is legal and the communication continues.

(5) According to $P_2'$ , Alice measures the corresponding
particles in Bell basis and sends the measurement results (denoted
as $R_2'$ ) to Bob through the classical channel.

(6) Bob compares $R_2$  with $R_2'$ . If they are identical, Bob
considers Alice is legal. Otherwise, he stops this communication.

4. Obtain the key. Bob makes Bell measurements on his left
particles by twos. It should be emphasized that each pair he
measures is selected out randomly. Bob records the sequence
numbers of all the pairs and sends the record to Alice. Alice then
measures her corresponding particles in Bell basis. As discussed
in the above paragraphs, their measurement results would be
identical. Subsequently, Alice and Bob can obtain the key from
these results. For example, $|\Phi^+\rangle$, $|\Phi^-\rangle$,
$|\Psi^+\rangle$ and $|\Psi^-\rangle$ are encoded into $00$, $01$,
$10$ and $11$, respectively.

5. Renew $ID$. Alice and Bob cut a little part from the key to
``refuel'' the shared information $ID$ and the previous one is
discarded.

Thus the whole QKD and identification protocol is finished. By
this process, Alice and Bob can not only get secure key but also
identify each other.
\section{Security}
The above scheme can be regarded as secure. The reasons are as
follows:

1. Each user's identity is authenticated and it is impossible for
Eve to impersonate Alice (or Bob) and distribute key with Bob (or
Alice). In this protocol the shared information $ID$ is only known
to Alice and Bob. If Eve wants to impersonate Alice, she can not
decrypt $y$ correctly and get $P_2$  in step.3. Consequently Eve
can not present $R_2'$  with which Bob is satisfied. On the
contrary, if Eve wants to impersonate Bob, she will be detected,
too. Because Eve can not give such a ``ciphertext'' that Alice
would obtain suited $P_1$  and $R_1$ after she used $ID$ to
``decrypt'' this ``ciphertext''. Furthermore, Eve can not obtain
$ID$ from the qubits and the classical information transmitted.
Because $ID$ is used as a key of the one-time-pad cipher to
encrypt some random bit strings including $P_1$, $R_1$ and $P_2$ ,
Eve can not extract any information about $ID$, even through
repeated attempts. In addition, it makes the protocol more secure
that Alice and Bob would renew $ID$ when they get the key.

2. The key distributed can not be eavesdropped imperceptively.
There are two general eavesdropping strategies for Eve. One is
called ``intercept and resend'', that is, Eve intercepts the legal
particles and replaces them by her counterfeit ones. For example,
Eve generates the same EPR pairs and sends one particle from each
pair to Bob, thus she can judge Bob's measurement results as Alice
does in step.4. But in this case there are no correlations between
Alice's particles and the counterfeit ones. Alice and Bob will get
random measurement results when they detect eavesdropping in
step.2. Suppose both Alice and Bob use $s$ pairs particles to
detect eavesdropping, the probability with which they obtain the
same results is only $(1/4)^s$ . That is, Eve will be detected
with high probability when $s$ is big enough. The second strategy
for Eve is to entangle an ancilla with the two-particle state that
Alice and Bob are using. At some later time she can measure the
ancilla to gain information about the measurement results of Bob.
This kind of attack seems to be stronger than the first strategy.
However, we can prove that it is invalid to our protocol as
follows.

Because each particle transmitted in the channel is in a maximal
mixed state, there are no differences among all these particles
for Eve. Furthermore, Eve does not know Bob will put which two
particles together to make a Bell measurement. As a result, what
she can do is to make the same operation on each particle
transmitted. Let $|\varphi\rangle_{ABE}$  denote the state of the
two particles and the ancilla, where the subscripts $A$, $B$ and
$E$ express the particles belonging to Alice, Bob and Eve,
respectively. Note that we do not limit each ancilla's dimension,
and allow Eve to build all devices that are allowed by the laws of
quantum mechanics. What we wish to show is that if this
entanglement introduces no errors into the QKD procedure, then
$|\varphi\rangle_{ABE}$ must be a product of a two-particle state
and the ancilla. This implies that Eve will gain no information
about the key by observing the ancilla or, conversely, if Eve is
to gain information about the key, she must invariably introduce
errors.

Without loss of generality, suppose the Schmidt decomposition
\cite{QCQI} of $|\varphi\rangle_{ABE}$  is in the form

\begin{eqnarray}
|\varphi\rangle_{ABE}=a_1|\psi_1\rangle_{AB}|\phi_1\rangle_E
                     +a_2|\psi_2\rangle_{AB}|\phi_2\rangle_E\nonumber\\
                     +a_3|\psi_3\rangle_{AB}|\phi_3\rangle_E
                     +a_4|\psi_4\rangle_{AB}|\phi_4\rangle_E
\end{eqnarray}
where $|\psi_i\rangle$ and $|\phi_j\rangle$  are two sets of
orthonomal states, $a_k$ are non-negative real numbers
($i,j,k=1,2,3,4$ ).

Because $|\psi_i\rangle$ are two-particle (four-dimensional)
states, they can be written as linear combinations of
$|00\rangle$, $|01\rangle$, $|10\rangle$ and $|11\rangle$. Let
\begin{eqnarray}
|\psi_1\rangle=b_{11}|00\rangle+b_{12}|01\rangle+b_{13}|10\rangle+b_{14}|11\rangle\nonumber\\
|\psi_2\rangle=b_{21}|00\rangle+b_{22}|01\rangle+b_{23}|10\rangle+b_{24}|11\rangle\nonumber\\
|\psi_3\rangle=b_{31}|00\rangle+b_{32}|01\rangle+b_{33}|10\rangle+b_{34}|11\rangle\nonumber\\
|\psi_4\rangle=b_{41}|00\rangle+b_{42}|01\rangle+b_{43}|10\rangle+b_{44}|11\rangle
\end{eqnarray}
in which $b_{pq}$ ($p,q=1,2,3,4$) are complex numbers. Then
$|\varphi\rangle_{ABE}$ can be written, thanks to Eqs.(2) and (3),
as
\begin{widetext}
\begin{eqnarray}
|\varphi\rangle_{ABE}=|00\rangle_{AB}\otimes(a_1b_{11}|\phi_1\rangle+a_2b_{21}|\phi_2\rangle+a_3b_{31}|\phi_3\rangle+a_4b_{41}|\phi_4\rangle)_E\nonumber\\
                     +|01\rangle_{AB}\otimes(a_1b_{12}|\phi_1\rangle+a_2b_{22}|\phi_2\rangle+a_3b_{32}|\phi_3\rangle+a_4b_{42}|\phi_4\rangle)_E\nonumber\\
                     +|10\rangle_{AB}\otimes(a_1b_{13}|\phi_1\rangle+a_2b_{23}|\phi_2\rangle+a_3b_{33}|\phi_3\rangle+a_4b_{43}|\phi_4\rangle)_E\nonumber\\
                     +|11\rangle_{AB}\otimes(a_1b_{14}|\phi_1\rangle+a_2b_{24}|\phi_2\rangle+a_3b_{34}|\phi_3\rangle+a_4b_{44}|\phi_4\rangle)_E
\end{eqnarray}
\end{widetext}

For convenience, we define four vectors (not quantum states) as
follows:
\begin{equation}
v_l=(a_1b_{1l},a_2b_{2l},a_3b_{3l},a_4b_{4l}) \quad l=1,2,3,4
\end{equation}
Consider any two sets of particles on which Alice and Bob will do
ES, the state of the system is
$|\varphi\rangle_{ABE}\otimes|\varphi\rangle_{ABE}$. According to
the properties of ES, we can calculate the probability with which
each possible measurement-results-pair is obtained after Alice and
Bob measured their particles in Bell basis. For example, observe
the event that Alice gets $|\Phi^+\rangle$ and Bob gets
$|\Psi^+\rangle$ , which corresponds to the following item in the
expansion:
\begin{widetext}
\begin{equation}
\frac{1}{2}|\Phi^+\rangle_A|\Psi^+\rangle_B\otimes\left[\sum_{r,s=1}^4
(a_rb_{r1}a_sb_{s2}+a_rb_{r2}a_sb_{s1}+a_rb_{r3}a_sb_{s4}+a_rb_{r4}a_sb_{s3})
|\phi_r\phi_s\rangle_E\right]
\end{equation}

Therefore, this event occurs with the probability
\begin{equation}
P(\Phi_A^+\Psi_B^+)=\frac{1}{4}\sum_{r,s=1}^4|a_rb_{r1}a_sb_{s2}+a_rb_{r2}a_sb_{s1}+a_rb_{r3}a_sb_{s4}+a_rb_{r4}a_sb_{s3}|^2
\end{equation}
\end{widetext}
However, this event should not occur. In fact, if Eve wants to
escape from the detection of Alice and Bob, any results-pair other
than $\Phi^+\Phi^+$, $\Phi^-\Phi^-$, $\Psi^+\Psi^+$ and
$\Psi^-\Psi^-$ should not be appear. Let $P(\Phi_A^+\Psi_B^+)=0$ ,
we then have, from Eqs.(7) and (5),
\begin{equation}
v_1^Tv_2+v_2^Tv_1+v_3^Tv_4+v_4^Tv_3=0
\end{equation}
in which $v_l^T$ is the transpose of $v_l$.

Similarly, let the probabilities of $\Phi_A^+\Psi_B^-$,
$\Phi_A^-\Psi_B^+$ and $\Phi_A^-\Psi_B^-$ equal to 0, we get
\begin{equation}
v_1^Tv_2-v_2^Tv_1+v_3^Tv_4-v_4^Tv_3=0
\end{equation}
\begin{equation}
v_1^Tv_2+v_2^Tv_1-v_3^Tv_4-v_4^Tv_3=0
\end{equation}
\begin{equation}
v_1^Tv_2-v_2^Tv_1-v_3^Tv_4+v_4^Tv_3=0
\end{equation}
From Eqs.(8)-(11), we can obtain
\begin{equation}
v_1^Tv_2=v_2^Tv_1=v_3^Tv_4=v_4^Tv_3=0
\end{equation}
That is,
\begin{eqnarray}
\left\{\begin{array}{c}
v_1=0 \quad or \quad v_2=0\\
v_3=0 \quad or \quad v_4=0
\end{array}\right.
\end{eqnarray}
For the same reason, we can obtain the following results:\\
(1). Let the probabilities of $\Psi_A^+\Phi_B^+$,
$\Psi_A^+\Phi_B^-$, $\Psi_A^-\Phi_B^+$ and $\Psi_A^-\Phi_B^-$
equal to 0, we can get
\begin{eqnarray}
\left\{\begin{array}{c}
v_1=0 \quad or \quad v_3=0\\
v_2=0 \quad or \quad v_4=0
\end{array}\right.
\end{eqnarray}
(2). Let the probabilities of $\Phi_A^+\Phi_B^-$  and
$\Phi_A^-\Phi_B^+$ equal to 0, we then have
\begin{equation}
v_1^Tv_1-v_2^Tv_2+v_3^Tv_3-v_4^Tv_4=0
\end{equation}
\begin{equation}
v_1^Tv_1+v_2^Tv_2-v_3^Tv_3-v_4^Tv_4=0
\end{equation}
And then
\begin{eqnarray}
\left\{\begin{array}{c}
v_1=\pm v_4\\
v_2=\pm v_3
\end{array}\right.
\end{eqnarray}
(3). Let the probabilities of $\Psi_A^+\Psi_B^-$ and
$\Psi_A^-\Psi_B^+$ equal to 0, we can get the same conclusion as
Eq.(17).

Finally, we can obtain three results from Eqs.(13), (14) and (17):
\begin{description}
\item[\quad 1.] $v_1=v_2=v_3=v_4=0$ ; \item[\quad 2.]$v_1=v_4=0$
and $v_2=\pm v_3$; \item[\quad 3.]$v_2=v_3=0$ and $v_1=\pm v_4$
\end{description}
That is, each of these results makes Eve succeed in escaping the
detection of Alice and Bob. Now we can observe what the state
$|\varphi\rangle_{ABE}$ is by putting these results into Eq.(4).
If the first result holds, we have $|\varphi\rangle_{ABE}=0$,
which is meaningless for our analysis. Consider the condition
where the second result holds, $|\varphi\rangle_{ABE}$ can be
written as:
\begin{eqnarray}
|\varphi\rangle_{ABE}=(|01\rangle\pm|10\rangle)_{AB}\otimes(a_1b_{12}|\phi_1\rangle+\nonumber\\
a_2b_{22}|\phi_2\rangle+a_3b_{32}|\phi_3\rangle+a_4b_{42}|\phi_4\rangle)_E
\end{eqnarray}
It can be seen that $|\varphi\rangle_{ABE}$ is a product of a
two-particle state and the ancilla. That is, there is no
entanglement between Eve's ancilla and the legal particles, and
Eve can obtain no information about the key. Similarly, we can
draw the same conclusion when the third result holds.

To sum up, our protocol can resist the entangle-ancilla
eavesdropping strategy.
\section{Conclusion}
We have presented a QKD and identification protocol based on ES.
The security against the attack discussed in \cite{ZLG} is assured
by a classical means, ``randomly select the particles out and put
together by twos'', in stead of the quantum ones such as
alternative measurements \cite{ZYCP} or rotation of the Bell
states \cite{C2001,D2004,CQ-PH}. Furthermore, this classical means
brings us another advantage. That is, it is unnecessary to
randomize the initial Bell states as in \cite{C2000,C2001}. This
in turn leads to less Bell measurements in our protocol. For
instance, to distribute two bits of key, Alice and Bob make two
Bell measurements in our protocol, while in \cite{C2000,C2001}
they must make three. Therefore, we can draw a conclusion that
classical means is important to the research of quantum
cryptography and to some extent it is even more effective than the
quantum ones. Besides, classical means is easier to be
implemented. On the other hand, we have to confess that our
protocol has a disadvantage, i.e., it uses a sequence of entangled
states but not a single quantum system \cite{C2001,D2004,CQ-PH} to
generate the key. Fortunately, it is not a fatal problem. Many QKD
protocols work in this model, for example, the famous E91 protocol
\cite{E91}. Furthermore, each pair of particles is still in one of
the Bell states and can be reused in other applications after QKD.

\begin{acknowledgments}
This work is supported by the National Natural Science Foundation
of China, Grants No: 60373059; also supported by the National
Laboratory for Modern Communications Science Foundation of China,
Grants No: 51436020103DZ4001 and the ISN Open Foundation.
\end{acknowledgments}

\end{document}